\documentclass[preprint,3p,twocolumn]{elsarticle}
\biboptions{numbers,sort&compress}
\usepackage{amssymb}
\journal{Physics Lett A}
\usepackage{epsfig}
\usepackage{graphicx}
\usepackage{amsmath}
\usepackage{mathrsfs}
\usepackage{dcolumn}
\usepackage{ulem}
\newcommand{\Tr}{\text{Tr}}
\newcommand{\ket}[1]{|#1\rangle}

\begin{document}
\begin{frontmatter}

\title{Realization of quantum gates by Lyapunov control}

\author[]{S. C. Hou \corref{cor1}}
\ead{houshaocheng@mail.dlut.edu.cn}

\author[]{L. C. Wang}
\ead{wanglc@dlut.edu.cn}

\author[]{X. X. Yi}
\ead{yixx@dlut.edu.cn}

\cortext[cor1]{Corresponding author. Tel.: +86 13840934797}
\address{School of Physics and Optoelectronic Technology, Dalian
University of Technology, Dalian 116024, China}

\begin{abstract}
We propose a Lyapunov control design to achieve specific (or a family of)
unitary time-evolution operators, i.e., quantum gates in the Schr\"{o}dinger
picture by tracking control. Two examples are presented. In the first,
we illustrate how to realize the Hadamard gate in a single-qubit system, while in the second, the controlled-NOT (CNOT) gate is implemented in two-qubit systems with the Ising and Heisenberg interactions. Furthermore, we demonstrate that the control can drive the time-evolution operator into the local equivalence class of the CNOT gate and the operator keeps in this class forever with the existence of Ising coupling.
\end{abstract}


\end{frontmatter}

\section{Introduction}

Quantum information processing \cite{Nielsen} has become an
interdisciplinary research field covering the investigation of
fundamental questions in quantum physics \cite{kofler10},
metrology \cite{schmidt05,roos06} as well as the quest for a quantum
computer. The Hadamard gate and the Controlled-NOT (CNOT) gate are the
building blocks for quantum computing, which can be realized by
quantum control \cite{Alessandro,Dong}.
Various techniques have been developed for quantum
control \cite{Carlini,Palao,Lapert,Schirmer,Dong2,Wiseman,Verstraete,Sklarz},
which can be divided into coherent and in-coherent ones. Quantum
coherent optimal control is a powerful tool for designing control
fields, although finding the control fields is a time-consuming task.
Recently, it has found applications in many problems \cite{Carlini,Palao,Wu,XWang3,Rowland,Khaneja,Khaneja2,Fouquieres,Muller}.

Quantum Lyapunov control was proposed as a good candidate for quantum
state engineering. It has been well developed in both theory and
applications in the last decades \cite{Sklarz,Grivopoulos,Mirrahimi2,Kuang,XWang,WW,XWang2,WW2,Hou,Sharifi,Yang}.
The authors in Ref. \cite{Grivopoulos,Mirrahimi2,Kuang,XWang,Sharifi,Yang}
investigated different types of Lyapunov functions, schemes of field design and their convergence. In Ref. \cite{WW,XWang2,WW2},
Lyapunov control was applied to driving an open quantum system to its
decoherence-free subspace, preparing entanglement states, and
enforcing adiabatic evolutions.

In quantum Lyapunov control, the control fields are designed to make
the Lyapunov function decrease monotonically, while the system is
asymptotically steered to a desired state. The total Hamiltonian
of the system under control takes the form $H_0+\sum_n f_n(t)H_n$,
where $H_0$ is the free Hamiltonian which can usually not be turned off.
$H_n$ stand for the external control Hamiltonians with $f_n(t)$
the corresponding control fields. Lyapunov control can be
understood as a local optimization \cite{Sugawara} with the control
fields determined at every instant of time in feedback form.
Similar to quantum coherent optimal control, Lyapunov control
can be used to deal with different forms of Hamiltonian
systems. However, the calculation of control fields for Lyapunov control
is much easier since it does not need iteration. Another merit
of Lyapunov control is that the shape of control fields is
flexible \cite{Hou}.

Lyapunov control is mostly used to prepare quantum states. This method
can be extended to produce unitary operators (quantum gates) in view of
its advantages. Sklarz and Tannor studied the creation of quantum gates
in the subspace of a direct sum or direct product Hilbert space
by local-in-time control (Lyapunov control) working in the interaction
picture with respect to $H_0$ \cite{Sklarz}. However, in the presence
of the free Hamiltonian $H_0$, the time-evolution operator can usually
not reach a stationary one in the Schr\"{o}dinger picture. Our goal
in this letter is to prepare desired time-evolution operators in the
Schr\"{o}dinger picture by Lyapunov control such that quantum gates
might be easier to realize in the laboratory frame.
We use a tracking strategy for engineering time-evolution operators.
The Lyapunov function is designed to be a distance between the
time-evolution operator $U$ and $e^{-iH_0t}O$, where $O$ is the target
quantum gate. In this way, the control fields can steer $U$ to the orbit
(defined in Section 2) of $e^{-iH_0t}O$, such that the time-evolution
operator might reach $O$ at particular instances of time,
or stays in a desired family of operators forever.
Note that tracking the time-dependent operator $e^{-iH_0t}O$
in the Schr\"{o}dinger picture is equivalent to tracking $O$ in the
interaction picture. However, if $U_I(t)\rightarrow O$ in the interaction
frame, the evolution operator $U(t)$ in the Schr\"{o}dinger picture is
generally not $O$ and the gate time $t_g$ with $U(t_g)=O$ is not clear.
Therefore, it is more convenient to calculate the field in the
Schrodinger picture.

The letter is organized as follows. We present in Section 2
the Lyapunov function and the design of control fields based
on a tracking strategy. In Section 3.1, we demonstrate
the implementation of the Hadamard gate with a typical
single-qubit Hamiltonian. In Section 3.2, the CNOT
gate is realized in two-spin systems with the Ising and Heisenberg
interactions. We show further in Section 4 that the
time-evolution operator can stay in the local equivalence
class of the CNOT gate forever with the same control as in
Section 3.2. This is the main difference between the present
Lyapunov control and the other strategies discussed in the
literature \cite{Palao,Lapert,Khaneja,Khaneja2,Fouquieres,Muller}
where the aimed quantum gate is obtained at a fixed time.
Finally, we summarize our work in Section 5.

%---------------------section 2----------------------------

\section{General theory}
Our task is to design control fields to realize a target quantum gate
(or a desired family of quantum gates) by Lyapunov control in a closed system.
The dynamical equation for the time-evolution operator $U$ is
\begin{eqnarray}\label{eqn:schodinger}
i\frac{d U}{dt}=(H_0+\sum_n f_n(t)H_n)U,
\end{eqnarray}
where $H_0$ is the free Hamiltonian and $H_n$ $(n=1,2,3,...)$ are the
control Hamiltonians with $f_n(t)$ the control fields. At the initial time,
$U(0)=\text{I}$ where $\text{I}$ is the identity operator.
For simplicity, we set $\hbar=1$ throughout this letter. We restrict
our discussion to finite dimensional quantum systems where all the
operators can be represented by $N\times N$ matrices.

If the time-evolution operator $U$ is driven to $O$ (the target
operator) with all the control fields being turned off, $U$ will evolve as $U(t)=e^{-iH_0t}O$
which is usually different from $O$ at a later time.
To be specific, consider an $N$-dimensional system in the space spanned
by the eigenstates of $H_0$ which is written as $H_0=diag\{\lambda_1,\lambda_2,...\lambda_N\}$.
Then the time-evolution operator $U(0)=O$ governed by $H_0$ evolves as
\begin{eqnarray}
\hspace{-10mm}
U(t)&=&e^{-iH_0t}O \\
&=&\left(
\begin{array}{cccc}
 e^{-i\lambda_1t}O_{11}&e^{-i\lambda_1t}O_{12}&\cdots&e^{-i\lambda_1t}O_{1N} \\
 e^{-i\lambda_2t}O_{21}&e^{-i\lambda_2t}O_{12}&\cdots&e^{-i\lambda_2t}O_{2N} \\
  & &\cdots & \\
  e^{-i\lambda_Nt}O_{N1}&e^{-i\lambda_Nt}O_{N2}&\cdots&e^{-i\lambda_Nt}O_{NN} \nonumber\\
  \end{array}
\right),
\end{eqnarray}
where $O_{ij}$ $(i,j=1,2\cdots N)$ are the elements of the unitary
operator $O$. Clearly $U(t)$ evolves under $H_0$ except
the trivial case, $\lambda_1=\lambda_2=...=\lambda_N$. Therefore,
one can not asymptotically steer the system to a target operator $O$
as $t\rightarrow\infty$. Instead, we use a tracking strategy to
steer the time-evolution operator $U$ to track the evolving operator
\begin{eqnarray}
\tilde{O}(t)=e^{-iH_0t}O.
\end{eqnarray}
Once $U\rightarrow\tilde{O}(t)$ or evolves into the orbit of
$\tilde{O}(t)$, $U$ is expected to reach $O$ at later
times. Here, the orbit of $\tilde{O}(t)$ is defined as a set of
operators, $S=\{U|U=e^{-iH_0 T}\tilde{O}(t),
T\in \mathbf{R}\}=\{U|U=e^{-iH_0T}O,T\in \mathbf{R}\}$.
It is evident that the orbit of $\tilde{O}(t)$ is also the orbit of $O$.

The Lyapunov function here is based on the fidelity of two unitary
matrices
\begin{eqnarray}
F=\frac{|\Tr(U_1^{\dag}U_2)|}{N}, \label{eqn:Fid}
\end{eqnarray}
where $N=\Tr(U_1^{\dag}U_1)$ is the dimension of the system.
The fidelity is often used to measure the difference between two unitary
operators \cite{Palao,Ashhab,Thomas}. If $F=1$, $U_1$ and $U_2$ are
equal up to a non-physical global phase. With these notations, we
define the Lyapunov function as
\begin{eqnarray}
V=1-\frac{1}{N^2}|\Tr(\tilde{O}^{\dag}(t)U)|^2, \label{eqn:Lya}
\end{eqnarray}
which can be understood as a distance between $U$ and
$\tilde{O}(t)$ \cite{WangG}. The function satisfies $0\leq V\leq1$.
If $U=e^{i\theta}\tilde{O}(t)$ $(\theta\in \mathbf{R})$, then $V=0$.

In order to determine the control fields, we calculate the time
derivative of $V$,
\begin{eqnarray}
\dot{V}&=&-\frac{1}{N^2}\frac{d}{dt}\{\Tr(O^{\dag}e^{iH_0t}U)
[\Tr(O^{\dag}e^{iH_0t}U)]^*\}\nonumber\\
&=&-\frac{1}{N^2}\{\frac{d}{dt}\Tr(O^{\dag}e^{iH_0t}U)\
[\Tr(O^{\dag}e^{iH_0t}U)]^*\nonumber\\
&&+\Tr(O^{\dag}e^{iH_0t}U)\frac{d}{dt}[\Tr(O^{\dag}e^{iH_0t}U)]^*\}\nonumber\\
&=&-\frac{2}{N^2}\sum_n f_n(t)\Re\{\Tr(-iO^\dag e^{iH_0t}H_n U)\nonumber\\
&\cdot&[\Tr(O^\dag e^{iH_0t}U)]^*\},
\end{eqnarray}
where $\Re(...)$ denotes the real part of $(...)$. If we choose
\begin{eqnarray}
\hspace{-8mm}
f_n(t)&=&K\Re\{\Tr(-iO^\dag e^{iH_0t}H_n U)\ [\Tr(O^\dag e^{iH_0t}U)]^*\}\nonumber\\
&=& {K\Re\{\Tr(-i\tilde{O}^\dag(t) H_n U)[\Tr(\tilde{O}^\dag(t)
U)]^*}\}, \label{eqn:field}
\end{eqnarray}
where $K$ is a real positive number characterizing the strength
of control fields, we have $\dot{V}\propto-\sum_n f_n^2(t)\leq 0$,
enforcing a monotonic decrease of the Lyapunov function.
A time-dependent $K$ could be used as an envelope function
to modulate the amplitude of the control fields. For example, $K$
can be designed to avoid non-zero field at $t=0$ in order to be
experimentally feasible. Constant $K$ is adopted in this letter
for simplicity.

Tracking control can be classified by its goals into two categories:
trajectory tracking and orbit tracking \cite{XWang}. We wish to steer
the time-evolution operator $U$ into $\tilde{O}(t)$ or
into the orbit of $\tilde{O}(t)$ in order to reach the target $O$.
Our Lyapunov function is formally designed in the same way as in the
trajectory tracking, i.e, $U\rightarrow\tilde{O}(t)$ with $V\rightarrow 0$.
However, it is interesting that even if $V$ does not decrease to $0$, such a
design is still possible to steer $U$ to the orbit of
$\tilde{O}(t)$. To show this point, we define $U'=e^{-iH_0t'}O$,
which means $U'$ and $\tilde{O}(t)$ share the same orbit. The
Lyapunov function yields,
\begin{eqnarray}
V&=&1-\frac{1}{N^2}|\Tr[\tilde{O}^{\dag}(t)U']|^2\nonumber\\
&=&1-\frac{1}{N^2}|\Tr(O^{\dag}e^{iH_0t}e^{-iH_0t'}O)|^2\nonumber\\
&=&1-\frac{1}{N^2}|\Tr(e^{iH_0\Delta t})|^2\nonumber\\
&=&1-\frac{1}{N^2}|\sum_{n=1}^{N}{e^{i\lambda_n \Delta t}}|^2\nonumber\\
&\geq&1-\frac{1}{N^2}\left(\sum_{n=1}^{N}{|e^{i\lambda_n \Delta t}|}\right)^2=0
\end{eqnarray}
where $\lambda_i$ is one of the eigenvalues of $H_0$ and $\Delta
t=t-t'.$ Note that $\lambda_n$ is real since $H_0$ is Hermitian, so
$|e^{i\lambda_n \Delta t}|=1$. This means even $V$ does not decrease
to $0$, it is still possible to produce the target unitary
time-evolution operator. The Lyapunov function might also be defined
as the minimum distance between $U$ and $S$ (the orbit of $\tilde{O}(t)$).
In this way, when $U$ is driven to $S$, $V\rightarrow0$. However,
this design will require more complicated calculations due to the
minimization of the distance. We'll show in the following sections
that the Lyapunov function Eq.(\ref{eqn:Lya}) is effective to produce
quantum gates.

We then explain how the target operator is obtained. With $\tilde{O}(0)=O$
in mind, when $\tilde{O}(t)$ has a good recurrence property, it is clear that $U$
will reach $O$ precisely at some finite times if $U$ is driven into the orbit of
$\tilde{O}(t)$. In the case of nonrecurrent $\tilde{O}(t)$, assume $U$ is driven
to the orbit of $\tilde{O}(t)$ at $t$. Then, $U(t)$ can be expressed as
$U(t)=\tilde{O}(t-t')$ because the Lyapunov function (distance of $U$ and
$\tilde{O}(t)$) does not need to be zero. If $t-t'=s<0$, we are sure that
after a finite time $-s$, $U(t+(-s))=\tilde{O}(0)=O$, the desired operator
can still be reached precisely. In addition, it is worth mentioning that the
tracked operator $\tilde{O}(t)$ is not unique in the sense that all operators
in the form of $\tilde{O}(t+\tau)$ ($\tau$ is a real constant) are
equivalent to $\tilde{O}(t)$. The parameter $\tau$ can be chosen freely
(positive or negative) making it easier to drive $U$ to $\tilde{O}(s)$ $(s<0)$
such that the desired gate can be reached precisely. It is easy to see that
the orbit of $\tilde{O}(t+\tau)$ is also $S$ according to our definition.

%-----------------------section3--------------------------------------

\section{Quantum gates by Lyapunov control}
In the circuit model of quantum computation, a quantum gate (or
quantum logic gate) is a basic quantum circuit operating on a small
number of qubits. It is the building block of a quantum computer,
like the classical logic gate for contemporary computers. It is proved
that any unitary transformation can be decomposed into single-qubit
and two-qubit gates \cite{Nielsen}. Thus the two kinds of gate play
a fundamental role in quantum computation. In the following, we
demonstrate how to use the Lyapunov method to achieve the
single-qubit Hadamard gate and the two-qubit CNOT gate.

\subsection{Single-qubit gates}
The Hadamard gate acts on a single qubit. It maps the basis state
$|0\rangle$ to $1/\sqrt{2}(|0\rangle+|1\rangle)$ and $|1\rangle$ to
$1/\sqrt{2}(|0\rangle-|1\rangle).$ This operation can be
represented by the following matrix,
\begin{eqnarray}
O_H=\frac{1}{\sqrt{2}}\left(
    \begin{array}{cc}
        1 & 1 \\
        1 & -1 \\
      \end{array}
    \right).
\end{eqnarray}
\begin{figure}
\includegraphics*[width=7.3cm]{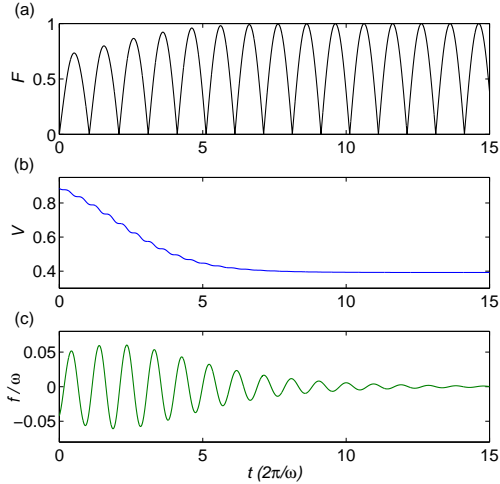}
\caption{(Color online) Time evolution of (a) the fidelity $F$, (b)
the Lyapunov function $V$ and (c) the control field $f(t)$ for the
Hadamard gate. The operator $U$ is driven to $O_H$ with $F\approx
0.999$ at $t=9.62$. As the Lyapunov function decrease monotonically,
fidelity larger than $0.999$ can be achieved at times later than
$t=9.62$ . $K=0.05\omega$ and $\tau=1/\omega$ is chosen in this figure.} \label{FIG:HDM}
\end{figure}
The Hamiltonian for a controlled single-qubit system can be
expressed as
\begin{eqnarray}
H=H_0+f(t)H_1
\end{eqnarray}
where $H_0=\frac{\omega}{2}\sigma_z$ is  the free Hamiltonian and
$H_1=\sigma_x$ represents the control Hamiltonian with $f(t)$ the
control field. Within the time scale where decoherence is
ignorable, the time-evolution operator $U$ is governed by
\begin{eqnarray}
i\frac{d U}{dt}=(H_0+f(t)H_1)U.
\end{eqnarray}

We now show how to realize the Hadamard gate $O_H$ by the Lyapunov
control. According to our theory, we can choose the Lyapunov function as
$V=1-\frac{1}{N^2}|\Tr[\tilde{O}_H^{\dag}(t)U]|^2$, where $N=2$ and
$\tilde{O}_H(t)=e^{-iH_0t}O_H$. However, this Lyapunov function leads
to $f(t)=0$ at the beginning of the control, an initial short
non-zero control field is thus required to trigger the control.
Alternatively, this problem can be solved by adopting
$\tilde{O}^{\dag}(t+\tau)$ instead of $\tilde{O}^{\dag}(t)$ in the
Lyapunov function as addressed in Section 2, i.e.,
\begin{eqnarray}
{V=1-\frac{1}{N^2}|\Tr[\tilde{O}_H^{\dag}(t+\tau)U]|^2}.
\end{eqnarray}
From this Lyapunov function the control field follows,
\begin{eqnarray}
\hspace{-10mm}
f(t)=K\Re\{\Tr(-i\tilde{O}^\dag_{H}(t\!+\!\tau) H_1 U)
[\Tr(\tilde{O}^\dag_{H}(t\!+\!\tau) U)]^*\}.
\label{eqn:fieldHDM}
\end{eqnarray}

Numerical simulation results are shown in  Fig.\ref{FIG:HDM},  where
we plot the fidelity between $U$ and $O_H$ (defined by
Eq.(\ref{eqn:Fid})), the Lyapunov function and the control field as a
function of time. The time-evolution operator $U$ reaches $O_H$ with
fidelity $F\approx0.999$ at $t=9.62$, see Fig.\ref{FIG:HDM} (a).
Although the Lyapunov function does not decrease to $0$ as shown in
Fig.\ref{FIG:HDM} (b), the time-evolution operator $U$ is
driven to the orbit of $\tilde{O}_H(t+\tau)$, and then arrives at
$O_H$ periodically with time.
In this model, the Hadamard gate with high fidelity
($F\rightarrow1$) can be achieved for any value of $K$ and $\tau$
in a finite time (except a $\tau$ that leads to $f(t)=0$).

Any single-qubit rotation can be expressed as
$U_R=e^{-i\frac{\theta}{2}\vec{n}\cdot \vec{\sigma}}$ where $\theta$
is the rotation angle around the axis $\vec{n}=(\sin\omega\cos\phi,\sin\omega\sin\phi,\cos\omega)$ in the
Bloch sphere. We further simulate our model with a large number of
different target quantum gate $U_R$. For each $U_R$ ,
$\omega$ (from $0$ to $\pi$), $\theta$ (from $0$ to $2\pi$) and
$\phi$ (from $0$ to $2\pi$) are randomly chosen. The results
suggest that the proposed technique can produce any single-qubit
rotations. The control mechanism of implementing a single-qubit gate $U=\left(\begin{array}{cc}
 a_0 & a_1 \\
b_0 & b_1 \\
\end{array}
\right)$ with the free Hamiltonian $H_0=\sigma_z$ is interpreted as follows. First, the
Lyapunov control (plus $H_0$) tips the basis state $\ket{0}$($\ket{1}$) to the latitude
on the Bloch sphere where
$\left[\begin{array}{cc}
a_0(a_1)\\
b_0(b_1)\\
\end{array}
\right]$ belongs. Then, free evolution (z-rotation) will drive the states to reach
$\left[\begin{array}{cc}
  a_0(a_1) \\
  b_0(b_1)\\
\end{array}
\right]$
periodically.

\subsection{Two-qubit gates}

The controlled-NOT (CNOT) gate is widely used in quantum
information processing, which flips the target qubit if and only if
the controlled qubit is in state $\ket{1}$. It together with arbitrary
single-qubit gates composes a set of universal quantum
gates, namely, any operation possible on a quantum computer can be
reduced as a finite sequence of gates from the universal gates
\cite{Nielsen}. The matrix representation for this gate in the bases
$\ket{00},\ket{01},\ket{10},\ket{11}$ is,
\begin{eqnarray}
O_{C}=\left(
    \begin{array}{cccc}
        1 & 0 & 0 & 0 \\
        0 & 1 & 0 & 0 \\
        0 & 0 & 0 & 1 \\
        0 & 0 & 1 & 0 \\
      \end{array}
    \right),
\end{eqnarray}
where the first qubit is the control qubit and the second is the
target qubit.

Consider an NMR system with two spins coupled via Ising
interaction, the Hamiltonian of such a system reads,
\begin{eqnarray}
H_0=\frac{\omega_1}{2}\sigma_{z}^{(1)}+\frac{\omega_2}{2}\sigma_{z}^{(2)}+
\frac{J}{4}\sigma_{z}^{(1)}\otimes\sigma_{z}^{(2)},
\label{eqn:Ising}
\end{eqnarray}
where $\omega_1$ and $\omega_2$ are the precession frequencies of the
two spins and $J$ represents the coupling strength. We show that
the CNOT gate can be realized by shinning a magnetic field on the
second spin (qubit) via the control Hamiltonian,
\begin{eqnarray}
H_1=\sigma_{x}^{(2)}.
\label{eqn:H1}
\end{eqnarray}
The time-evolution operator $U$ satisfies,
\begin{eqnarray}
i\frac{d U}{dt}=(H_0+f(t)H_1)U.
\label{eqn:evlcnot}
\end{eqnarray}
\begin{figure}
\includegraphics*[width=7.3cm]{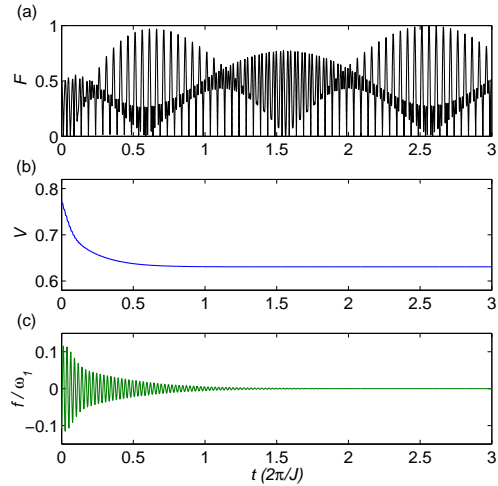}
\caption{(Color online) The implementation of the CNOT gate with Ising coupling.
The figure shows the (a) fidelity, (b) Lyapunov function and (c) control
field versus time $t$. The time-evolution operator $U$ is driven to
$O$ with $F \approx 0.9999$ at $t=2.56$ while the Lyapunov function
decrease monotonically, although not convergent to 0. The
parameters are, $\omega_2=2 \omega_1$, $J=0.05\omega_1$,
$K=0.1\omega_1$ and $\tau=0.3/\omega_1$} \label{FIG:CnotIsing}
\end{figure}
where the control field $f(t)$ can be realized by a time-dependent
magnetic field. In this example, we still use $\tilde{O}(t+\tau)$ to
define the Lyapunov function where different $\tau$ may lead to different
fidelity. This is different from last example where a
high fidelity (very close to 1) can be achieved for any $\tau$.

Now we go to the details. The Lyapunov function is defined as,
\begin{eqnarray}
V=1-\frac{1}{N^2}|\Tr(\tilde{O}^{\dag}_{C}(t+\tau)U)|^2,
\label{eqn:LyaIsing}
\end{eqnarray}
where $\tilde{O}_{C}(t+\tau)=e^{-iH_0(t+\tau)}{O}_{C}$ and $N=4$.
The control fields are given by,
\begin{eqnarray}
\hspace{-10mm}
f(t)=K\Re\{\Tr(-i\tilde{O}^\dag_{C}(t\!+\!\tau) H_1 U)
[\Tr(\tilde{O}^\dag_{C}(t\!+\!\tau) U)]^*\}.
\label{eqn:fieldIsing}
\end{eqnarray}

\begin{figure}
\includegraphics*[width=8cm]{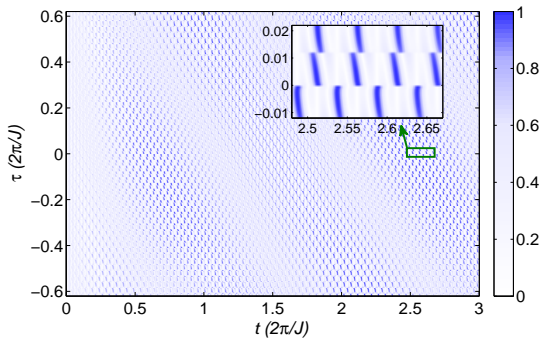}
\caption{(Color online) Fidelity versus $t$ and $\tau$. The
parameters chosen are the same as in Fig.\ref{FIG:CnotIsing}.
The characteristic time scale of $\tilde{O}(t+\tau)$
is $1/\omega_1$ and $1/J$ ($1/J \geq 1/ \omega_1$), thus the
range of $\tau$ is chosen with a scale of  $1/J$ and both
positive and negative $\tau$ is considered. With the help
of this figure, we can choose optimal $\tau$ and $t$ with
both high fidelity and short time for implementing the CNOT gate.}
\label{FIG:tau}
\end{figure}

We numerically simulate the model and plot the fidelity
$F=\frac{|\Tr(O_C^{\dag}U)|}{4}$, the Lyapunov function $V$ and the
control field $f(t)$ as as a function of time in
Fig.\ref{FIG:CnotIsing}. We find that the fidelity reaches about
$0.9999$ at $t=2.56$ as shown in Fig.\ref{FIG:CnotIsing} (a).
The parameter $\tau$ can be found numerically for a better
performance. For example, we plot the evolution of fidelity with
different $\tau$ in Fig.\ref{FIG:tau}. Such a figure shows appropriate
$\tau$ as well as the time $t$ to achieve the CNOT gate.
It is shown that the fidelity oscillates with $t$ which
originates from the free Hamiltonian and can not be eliminated in
the Schrodinger picture. Thus the gate time $t_g$ needs to be
precisely chosen for high fidelity. Since $f(t)\rightarrow0$ before
$t_g$, in this sense, the fidelity is robust against the switching
time of $f(t)$. For a given gate time $t$, the fidelity also depends
on $\tau$ and may change dramatically at certain values of $\tau$
(we call these values $\tau'$) as seen in Fig.3. However, when
implementing a quantum gate, $\tau$ and $f(t)$ are
known in advance by computer simulation. So the robustness
against $\tau$ need not be considered in 
experiments. The reason for the sudden change of fidelity is that
when $\tau=\tau'$, the tracked operator $\tilde{O}_C(t+\tau')$ is
ineffective which leads to $f(0)=0$ and  $f(t)\approx0$ when $t$
is small. Then, $\tau_1=\tau'+\varepsilon$ and
$\tau_2=\tau'-\varepsilon$ ($\varepsilon$ is infinitesimal)
will generate significantly different control fields as well
as fidelities although $\tau_1\approx\tau_2$. See that a
typical $\tau'$ is $0$.

\begin{figure}
\includegraphics*[width=7.3cm]{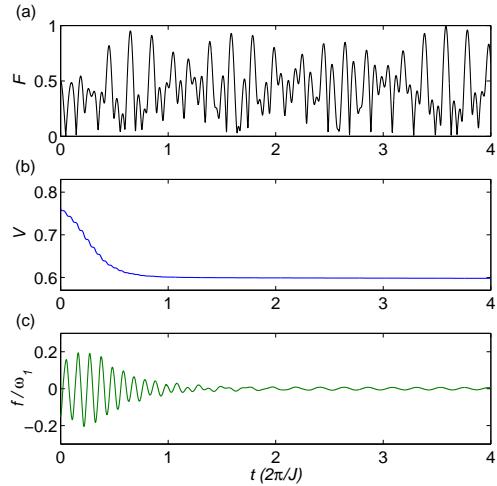}
\caption{(Color online) The implementation of the CNOT gate with Heisenberg
coupling. The figure shows (a) the fidelity, (b) the Lyapunov
function and (c) the control field as a function of $t$. Parameters
are set as $\omega_2=2\omega_1$, $J=0.2\omega_1$, $K=0.2\omega_1$
and $\tau=0.2/\omega_1$.  Note that despite
$\omega_2-\omega_1=5J$, and $\omega_2=2\omega_1=10J$, the Heisenberg
interaction can not be approximated by the Ising interaction, this
can be found by calculating the distance between the two
time-evolution operators, corresponding to the two interaction,
respectively.} \label{FIG:CnotHeis}
\end{figure}

The Ising coupling favors the implementation of the CNOT gate since
the time-evolution operator governed by $H=J\sigma_z\otimes\sigma_z$
is equivalent to a CNOT gate up to one-qubit rotations
\cite{Schuch}. In contrast, more operations are needed to
realize the CNOT gate with the Heisenberg coupling \cite{Schuch}.
Fortunately, the Heisenberg coupling may be reduced to the Ising
coupling when the condition $|\omega_1-\omega_2|\gg J$ is
satisfied \cite{Vandersypen,Zhang}.  We now show that the CNOT gate can
also be realized in a two-qubit system with Heisenberg coupling
in a similar manner as in the case of Ising coupling.

Consider a two-qubit system with Heisenberg coupling, the free
Hamiltonian takes,
\begin{eqnarray}
H_0=\frac{1}{2}\omega_1\sigma_{z}^{(1)}+\frac{1}{2}\omega_2\sigma_{z}^{(2)}+
\frac{1}{4}J\vec{\sigma}^{(1)}\cdot\vec{\sigma}^{(2)},
\label{eqn:Heis}
\end{eqnarray}
where
$\vec{\sigma}^{(j)}=(\sigma_x^{(j)},\sigma_y^{(j)},\sigma_z^{(j)}),
j=1,2$. With the same Lyapunov function Eq.(\ref{eqn:LyaIsing}), control
Hamiltonian Eq.(\ref{eqn:H1}) and design of control
field Eq.(\ref{eqn:fieldIsing}), we simulate the model and plot
the fidelity , Lyapunov function and control field as a function of time
in Fig.\ref{FIG:CnotHeis}. In this simulation, a stronger coupling
constant $J$ is used such that the Heisenberg interaction can not
be approximated by the Ising one. A fidelity about $0.994$ is
obtained at $t=3.58$. Further numerical simulations show that
for the Heisenberg Hamiltonian Eq.(\ref{eqn:Heis}) with strong
couplings, say $J\sim \omega_i, i=1,2$, the implementation of
the CNOT gate may have lower fidelity.

In this section, we use a single control field designed by the
Lyapunov method to realize the CNOT gate in two-qubit quantum
systems, different types of inter-qubit coupling are considered.
It is worth noting that in these examples, the control is also
effective if the control Hamiltonian Eq.(\ref{eqn:H1}) is replaced
by $H_1=\sigma_x^{(1)}+\sigma_x^{(2)}$, this indicates that our
proposal applies to homonuclear systems where two spins are
coupled simultaneously to a single RF field, and the precession
frequencies $\omega_1$ and $\omega_2$ in spins are replaced by
the chemical shifts in the rotating frame \cite{Wu}.
This method can also be used to in NMR systems for other purposes
to reduce the steps of operations. For example, in a homonuclear
two-spin system with coupling, a shaped non-selective hard pulse
acting on both spins can perform a local quantum gate on spin 1
 while freezing spin 2 without refocusing technology \cite{Nielsen,Vandersypen,Zhang}.

\section{Local equivalence operators}
In the last section, we have shown that the Hadamard gate and the CNOT
gate can be implemented by Lyapunov control. The fidelity reaches
almost $100\%$ at specific times, but it would change
after the gate time. In this section, we will show that by our method,
the time-evolution operator for the models in Section 3.2
can be steered into a target class of operators (the local equivalence
class of the CNOT gate) and stays in this class
forever when the control
fields are turned off.

Two two-qubit unitary operators $U_1$ and $U_2$ are called locally
equivalent if they can be connected by local operations, i.e.,
$U_1=L_1U_2L_2$, where $L_1,L_2\in SU(2)\otimes SU(2)$ are
the combinations of single-qubit operations. Here we denote the local
equivalence class of a unitary operator $O$ as $[O]$. Usually the
realization of two-qubit gates are more costly (e.g., taking longer
time, requiring more operations and so on) than that of single-qubit gates,
hence a difficultly implemented two-qubit operation can be realized
through its equivalent gate. On the other hand, any entangling
two-qubit gate with single-qubit gates forms a universal set of
quantum gates for quantum computing. Therefore, it is interesting
to study how to realize the local equivalence gate for some particular
two-qubit gates \cite{Muller} such as the CNOT gate.

Makhlin proposed three local invariants \cite{Makhlin} to
characterize the non-local property of a two-qubit gate $U \in U(4)$,
they are $d_1=\text{Re} G_1,d_2=\text{Im} G_1,$ and $d_3=G_2$, where
$G_1=\Tr^2 m_U \det U^{\dag}/16$ is complex and $G_2=(\Tr^2
m_U-\Tr m_U^2)\det{U^{\dag}/4}$ is real. $m_U$ is defined as
$m_U=Q^{T}U^{T}Q^*Q^{\dag}UQ$ with
\begin{eqnarray}
Q=\frac{1}{\sqrt{2}}\left(
    \begin{array}{cccc}
        1 & 0 & 0 & i \\
        0 & i & 1 & 0 \\
        0 & i & -1 & 0 \\
        1 & 0 & 0 & -i \\
      \end{array}
    \right).
\end{eqnarray}
Two two-qubit unitary gates are locally equivalent if they have the
same $d_i$ $(i=1,2,3)$. In order to quantify the distance between a
unitary operator $U$ and the equivalence set of the CNOT gate
$[O_C]$, we define
$D=\sqrt{(d_1-d'_1)^2+(d_2-d'_2)^2+(d_3-d'_3)^2}$ as a measure,
where $d_i$ and $d'_i$ are the invariants of $U$ and $O_C$,
respectively.

\begin{figure}
\includegraphics*[width=7.3cm]{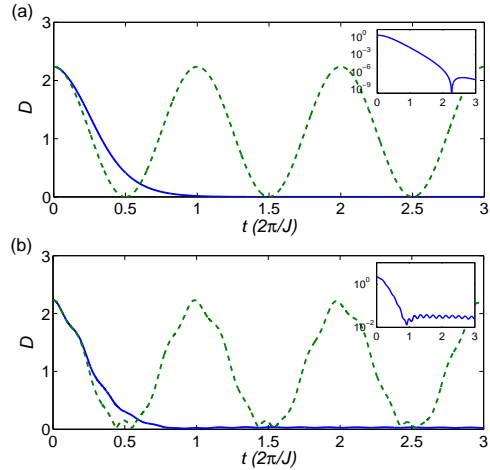}
\caption{(Color online) Driving the time-evolution operator to local
equivalence class of CNOT gate with (a) the Ising interaction and
(b) the Heisenberg interaction. Blue solid lines represent
time-dependence of $D$ under control (insets shows $D$ on a
logarithmic scale). Green dashed lines correspond to $D$ without
control ($f(t)=0$). The parameters used for (a) and (b) are the same
as that in Fig. \ref{FIG:CnotIsing} and Fig.\ref{FIG:CnotHeis},
respectively.} \label{FIG:eqcnot}
\end{figure}

For the Ising model Eq.(\ref{eqn:Ising}), we find that once the time
evolution operator is driven to the orbit of $\tilde{O}_{C}(t+\tau)$, it
would stay in $[O_C]$ forever, even if the control fields are turned off.
Now we show this point in detail. For the Ising model without control,
the time-evolution operator in the orbit of $\tilde{O}_{C}(t+\tau)$
satisfies $U=e^{i\theta}e^{-iH_0t'}O_C$, where $\theta$ and $t'$ are
 real numbers. $U$ can be decomposed as $U=e^{i\theta}U_aU_bO_C$,
 where $U_a=e^{-i(\frac{\omega_1}{2}\sigma^{(1)}_z+\frac{\omega_2}{2}\sigma^{(2)}_z)t'}$
(local operation) and $U_b=e^{-i\frac{J}{4}\sigma^{(1)}_z\otimes\sigma^{(2)}_zt'}$. If
the distance $D$ between $M=e^{i\theta}U_bO_C$ and $O_C$ is zero,
then $M$ and $O_C$ are locally equivalent. To calculate the distance,
we need $m_M$ that takes,
\begin{eqnarray}
m_M=Q^{T}M^{T}Q^*Q^{\dag}MQ\quad\quad\quad\quad\quad\nonumber\\
=e^{i2\theta}\!\left(
    \begin{array}{cccc}
        0 \quad\ \ \ \sin(\frac{Jt'}{2}) \quad -\cos(\frac{Jt'}{2})\quad\quad  0\\
        \sin(\frac{Jt'}{2})\quad\ \ \ \ 0\quad \quad 0 \quad\ \  -\cos(\frac{Jt'}{2})\\
        -\cos(\frac{Jt'}{2})\quad\ 0\quad \quad 0\quad\ \ -\sin(\frac{Jt'}{2}) \\
        \ 0 \quad -\cos(\frac{Jt'}{2}) \quad -\sin(\frac{Jt'}{2})\quad \quad 0\\
      \end{array}
    \right).
\label{eqn:mmatrix}
\end{eqnarray}
With the definition of local invariants, we have $d_1=\Re G_1=0,
d_2=\Im G_1=0$ and $d_3=G_2=1.$ which are also the local invariants
of $O_C$ regardless of $t'$, $J$ and $\theta$. Therefore, $U$ and $O_C$
are locally equivalent.
This fact can be exploited to create the CNOT gate at a more
flexible time. Note that when the coupling $\sigma_z^{(1)}\sigma_z^{(2)}$
can not be switched off, the CNOT gate can be achieved only at specific
times with the help of local operations \cite{Schuch}. However, if the
evolution operator keeps in $[O_C]$, the CNOT gate can be obtained at
any time with the help of local operations.

Next, we numerically simulate Eq.(\ref{eqn:evlcnot}) with the same
parameters ($\omega_{1,2}, J, K$ and $\tau$ ) as in
Fig.\ref{FIG:CnotIsing} and calculate the distance between
the time-evolution operator $U$ and $[O_C]$. The result is
plotted with blue solid line in Fig.\ref{FIG:eqcnot} (a).
We find that the time-evolution operator is driven to $[O_C]$
with high precision and stays in $[O_C]$ forever.
Note that without the Lyapunov controls, the local invariants of $U(t)$
(obtained by setting $H_1=0$ in Eq.(\ref{eqn:evlcnot}) with initial
condition $U(0)=I$) evolve as $d_1=\cos^2(\frac{J}{2}t),d_2=0,d_3=2+\cos(Jt)$.
The distance between $U(t)$ and $[O_C]$ is $D=\sqrt{5}\cos^2(\frac{J}{2}t)$,
which reaches zero only when $t=(2n+1)\pi/J$ shown by the green-dashed
line in Fig.\ref{FIG:eqcnot}(a). Simulations with different $\omega_1$,
$\omega_2$ and $J$ suggest that the time-evolution operator can be
driven to $[O_C]$ for a wide range of parameters.

For the two-spin model with the Heisenberg interaction Eq.(\ref{eqn:Heis}),
the time-evolution operator $U=e^{i\theta}e^{-iH_0t'}O_C$ can not stay in $[O_C]$.
Nevertheless, it is still possible to
drive the time-evolution operator approximately to $[O_C]$
when the two spins are weakly coupled and have a large difference
at the precession frequency. To illustrate this, we simulate
the Heisenberg model with the same parameters as in Fig.\ref{FIG:CnotHeis}
and plot the distance $D$ between $U$ and $[O_C]$ in
Fig.\ref{FIG:eqcnot}(b) (blue solid line). We find that
the time-evolution operator can be driven to $[O_C]$, but the
performance is not as perfect as that with the Ising interaction.
Large difference at the precession frequencies and weaker
coupling between the spins can improve the performance.

\section{Summary}

We present a Lyapunov control design to produce a quantum gate
(or a class of quantum gates) in the Schr\"{o}dinger picture.
Considering that a unitary operator is usually not stationary
under free evolution, a tracking strategy is adopted to steer the
time-evolution operator to the orbit of target operator so as to
reach the target. We introduce an adjustable parameter $\tau$ into
the tracked operator such that the time-evolution operator can be
easily driven to the target operator with high precision.
We apply the proposal to the implementation of the Hadamard gate and
the CNOT gate at some instance of time. Besides, we find that with
the Ising interaction, the time-evolution operator can be driven into
the local equivalence class of the CNOT gate and stay in that class forever.
The advantages of the traditional Lyapunov control, e.g., easy and
flexible design of control fields, no measurement induced decoherence,
will contribute to the implementation of quantum gates.
Meanwhile, there are some limitations to be improved in our method.
First, the gate time $t_g$ is determined after the simulation
which may be inconvenient when the control is applied. Second,
our method may not be effective in general to implement other
quantum gates or realize quantum gates in other (complex)
Hamiltonian systems. At last, the fact that the evolution
operator can be driven into the equivalence class of the CNOT
gate may not be available with other coupling Hamiltonians.

\section{Acknowledgements}
This work is supported by the NSF of China under Grants Nos
61078011, 10935010 and 11175032.

 \end{document}